\documentclass[preprint,12pt]{elsarticle}
\usepackage{amssymb}
\usepackage[]{epsfig}
\usepackage{amsmath}
\usepackage{pstricks}
\usepackage{graphicx}
\usepackage{fancybox}
\journal{Physics Letters B}
\begin{document}
\begin{frontmatter}

\title{The Cornell black hole}
\author{Euro Spallucci\footnote{e-mail:Euro.Spallucci@ts.infn.it} $\,$\footnote{Senior Associate}}
\address{INFN,\\
Sezione di Trieste, Trieste, Italy}
\author{Anais Smailagic\footnote{e-mail:Anais.Smailagic@ts.infn.it} $\,$\footnote{Senior Associate}}
\address{INFN, Sezione di Trieste, Italy}

\begin{abstract}
The Cornell potential can be derived from a recently proposed non-local extension 
of Abelian electrodynamics. Non-locality can be alternatively described by an extended charge distributions in Maxwell electrodynamics. We state that in these  models the energy momentum tensor necessarily requires the presence of the interaction term between the field and the charge itself. We show that this extended form of energy momentum tensor leads
to an exact solution of the Einstein equations describing a charged AdS black hole. We refer to it as
the "~Cornell black hole~"(CBH).  Identifying
the effective cosmological constant with the pressure of Van der Waals fluid, we study the gas-liquid phase transition and determine the critical parameters.
\end{abstract}
\end{frontmatter}

\section{Introduction}
The Cornell potential is a phenomenological  potential exhibiting long distance confinement.
 It was originally introduced  to reproduce charmonium spectrum \cite{Eichten:1978tg}.  
 To obtain such a  potential  out of a  Yang-Mills gauge theory is still challenging since perturbation approach is not applicable in the strong coupling regime. On the other hand, it has long been
 acknowledged  that confinement is, in fact, an "~Abelian~" long distance phenomenon \cite{Luscher:1978rn}. However, obtaining an effective abelian approximation of $QCD$ turns out to be a lengthy and non-trivial task \cite{Kondo:1997pc,Kondo:1997kn,Kondo:2014sta}. \\
 It has been recently noticed that, for a special class of null $SU(N)$ gauge potentials $A^a{}_\mu = c^a \phi(x)l_\mu$, where $l_\mu l^\mu=0$ and $c^a$ is constant vector in color space, Yang-mills field equations reduce to a system of decoupled  Maxwell equations. This particular choice for the gauge field allows to establish an intriguing \emph{duality} between Yang-Mills theory and  gravity ( see \cite{Bern:2019prr} for a general review of this topic.). As an outcome,  the dual of the Coulomb potential turns out to be  a Schwarzschild black hole.
 \\
 Following this line of reasoning,  we recently described  the Cornell potential as a static solution of the field equations  of a non-local Abelian  gauge theory\cite{Smailagic:2020kep,Smailagic:2021poa}. We shall refer to this Lagrangian model  as \emph{Cornell electrodynamics}(CE).\\
 The  null Yang-Mills  fields, quoted above, are the electromagnetic  analogue of the Kerr-Schild decomposition of the metric tensor given by
 \begin{equation}
g_{\mu\nu} = \eta_{\mu\nu} -2\phi\left(\,x\,\right) l_\mu l_\nu 
\end{equation}
where, the null condition $l_\mu l^\mu=0$ holds  both with respect to Minkowski and the complete
metric tensor. This analogy is referred to as "~Kerr-Schild double copy~"  \cite{Monteiro:2014cda,Monteiro:2015bna,White:2017mwc,Bahjat-Abbas:2017htu,Spallucci:2022xtr}.
\\
The advantage of Kerr-Schild metric is to reduce  Einstein field equations   to a single  Poisson equation for the unknown scalar function $\phi\left(\,x\,\right)$:

\begin{equation}
R^0{}_0\equiv\nabla^2 \phi = -8\pi G_N \left(\, T^0{}_0 - \frac{1}{2} T^\mu{}_\mu\,\right) \ .
\label{Poi}
\end{equation}

The physical meaning of $\phi\left(\,x\,\right)$ as the relativistic gravitational potential becomes clear once the metric is written in the  spherical gauge:

\begin{equation}
ds^2= -\left(\, 1 + 2\phi(r)\,\right) dt^2 +\left(\, 1 + 2\phi(r)\,\right)^{-1}dr^2 +r^2\left(\, d\theta^2+\sin^2\theta d\varphi^2\,\right)\ .
\end{equation}

 In summary, once $T^{\mu\nu}$ is chosen, finding the corresponding curved metric is reduced to solving equation (\ref{Poi}) in Minkowski space.\\
When the electromagnetic part of energy-momentum tensor $T^{\mu\nu}$ is taken into account in the Einstein equation (\ref{Poi}) the result is proportional to $\nabla^2 \left(\, A^0\,\right)^2$, where $A^0$ is the time-like component of the gauge potential. Therefore, we conclude that the electromagnetic contribution to the metric tensor is of the form  $ A_0^2 $. For example, in the case of the Einstein equations coupled to a  static, point-like, Coulomb potential, one finds the usual Reissner-Nordström form of metric:

\begin{equation}
\phi_{RN}\left(\, r \,\right)=\phi_m +2\pi G_N \, A_0{}^2\ ,\qquad A_0=\frac{q}{4\pi  r}
\label{bhc}
\end{equation}
where $q$ is the electric charge. $\phi_m $ is the contribution from the mass, 
i.e. the Schwarzild term.\\

Our purpose in this Letter is to solve Einstein equations coupled to Cornell electrodynamics and obtain the Kerr-Schild double copy  of the confinement  potential. We expect the metric to keep the form 
of (\ref{bhc}), but with the Coulomb potential replaced by the Cornell one.\\\\
The paper is organized as follows. In Sect.(\ref{cef}) we discuss the equivalence between non-local
Cornell electrodynamics and Maxwell gauge theory with an extended source. The energy momentum tensor, including the interaction energy between the field and the extended charge, is described. Then, we solve the Einstein equations and find the Kerr-Schild double copy of the Cornell potential, i.e. 
the Cornell black hole. In Sect.(\ref{termo})  the thermodynamic properties of the CBH are described.  Identifying the effective cosmological constant with the pressure of Van der Waals fluid
we study the \emph{gas-liquid} phase transition and determine the critical parameters in terms of the
 mass and charge in the Cornell effective Lagrangian.
 In Sect.(\ref{end}) we summarize the main results obtained. Finally, in the Appendix we present 
 the computation of the energy momentum tensor part coming from the interaction term $J^\mu A_\mu$.

\section{Cornell effective Lagrangian.}
\label{cef}
The phenomenological potential between a quark anti-quark pair is  describes by the Cornell formula
\begin{eqnarray}
&& V_C\left(\, r\,\right)=-\frac{4}{3}\frac{\alpha_s}{r}+\sigma\,r \ ,\label{linconf}\\
&& \alpha_s\left(\, \mu^2\,\right)\equiv \frac{4\pi}{\left(\, 11-\frac{2}{3}n_f\,\right)\ln\left(\mu^2/\Lambda_{QCD}\right)}\ .
\end{eqnarray}
where
$ \alpha_s\left(\, \mu^2\,\right) $ is the strong running coupling constant;
$\sigma$ is the "~string tension~" between a quark anti-quark pair and the renormalization scale $\mu$ is chosen to be
\begin{equation}
\mu \equiv\frac{2m_Q m_{\bar{Q}}}{m_Q +m_{\bar{Q}}}
\end{equation}
with $m_Q$ and $m_{\bar{Q}} $  the masses of quark anti-quark, respectively. Finally, $n_f$ is the number of quark flavors and $\Lambda_{QCD}\simeq 0.15 GeV $ is the QCD energy scale.\\\\
In a recent paper we introduced a novel way to generate a confining linear potential in the framework of an Abelian gauge theory. Since the gauge potential (\ref{linconf}) is a sum of the Coulomb  and a linear term one has to modify  Maxwell electrodynamics by adding an inverse Lee-Wick term \cite{Smailagic:2020kep}. The result is the non-local Lagrangian
\begin{equation}
L\left[\, A\,\right] =
-\frac{1}{4}\,F_{\mu\nu}\, \frac{\partial^2}{\partial^2 - \Lambda_{QCD}^2}\,F^{\mu\nu}-g \, J^\mu \, A_\mu  \ .
 \label{lcornell1}
\end{equation}
$g$ is the usual gauge coupling constant. 
\\
Variation of (\ref{lcornell1}) with respect  $A_\mu$ gives the field equations    

\begin{equation}
\partial_\lambda\frac{\partial^2}{\partial^2-\Lambda_{QCD}^2}F^{\lambda\mu}=gJ^\mu\ .
\label{nlmax}
\end{equation}
(\ref{nlmax}) are a non-local version of Maxwell equations.\\
If  $J^\mu=\delta^\mu{}_0 \, J^0=\delta^\mu{}_0 \, \delta^{(3)}\left(\, \vec{x}\,\right)$, one obtains
\begin{eqnarray}
A^0&=&-\frac{g}{4\pi\, r} +\frac{g}{8\pi}\Lambda_{QCD}^2\, r \label{c1}\\
E&\equiv & -\frac{dA^0}{dr}=- \frac{g}{4\pi\, r^2} -\frac{g}{8\pi}\Lambda_{QCD}^2\nonumber
\end{eqnarray}
By introducing the following definitions
\begin{eqnarray}
&&\frac{g}{4\pi}=\frac{4}{3}\alpha_s\ ,\\
&&\frac{g}{8\pi}\Lambda_{QCD}^2=\sigma\ ,
\end{eqnarray}
one reproduces the Cornell potential (\ref{c1}).\\\\
The solution (\ref{c1}) can be also recovered from the \emph{equivalent Lagrangian}

\begin{equation}
\widetilde{L}\left[\, A\,\right] =
-\frac{1}{4}\,F_{\mu\nu}\,F^{\mu\nu}-g \widetilde{J}^\mu \, A_\mu  \ ,\qquad
\widetilde{J}^\mu\equiv \left(\, 1-\frac{\Lambda_{QCD}^2}{\partial^2}\,\right)\, J^\mu\ .
 \label{lcornell}
\end{equation}

The new version of the model turns out to be ordinary  Maxwell electrodynamics in the presence of a non-local current
$\widetilde{J}^\mu$. Thus, the original  point-like charge is replaced by  a non-local charge density.\\
 As already mentioned in the introduction, in the case of distributed charges the interaction energy between the field and its source cannot neglected.  It must be included in the Lagrangian.
Following the procedure discussed in the Appendix, one finds the following energy-momentum tensor:

\begin{equation}
T_{\mu\nu}= F_{\mu\alpha}F_\nu{}^\alpha -\frac{1}{4} g_{\mu\nu} F^2+\frac{g }{2}\, \widetilde{J}_{(\mu} A_{\nu)}\ .
\end{equation}

Notice that $T_{\mu\nu}$ is no more traceless due to the presence of an extended source 
\begin{equation}
T^\nu{}_{\nu}= g \, \widetilde{J}^\mu A_\mu\ .
\end{equation}

In the electrostatic case $F^{0m}\equiv E^m=-\partial^m A^0$. The source in the Poisson equation (\ref{Poi}) reads
\begin{eqnarray}
T^0{}_0 -\frac{1}{2}T^\nu{}_{\nu}
&&=-\frac{1}{2}\,\vec{E}{}^2 +\frac{g}{2}\, \widetilde{J}{}^0 A^0\ ,\nonumber\\
&&=-\frac{1}{2}\left(\,\nabla A^0\cdot\nabla A^0 -g\, \widetilde{J}^0 A^0\,\right)\ .
\label{T}
\end{eqnarray}
It is appropriate to recall that in classical electrodynamics  $\vec{E}^2/2$ is the field "~kinetic term~"
while $e J^0 A_0/2$  represents the "~interaction~" energy between the field and the source.\\
The final step is to rewrite (\ref{T}) in a convenient way:

\begin{equation}
\nabla A^0\cdot\nabla A^0 -g\, \widetilde{J}^0 A^0 =\nabla A^0\cdot\nabla A^0 -A^0 \nabla\vec{E}=\frac{1}{2} \nabla^2\,A_0^2
\end{equation}

Thus, the Poisson equation (\ref{Poi}) has a general solution of the form

\begin{equation}
 \phi =\phi_0 +2\pi G_N\, A_0{}^2\ . \label{sol}
\end{equation}
$\phi_0=c_0 +c_1/r$ is the solution of the homogeneous equation $\displaystyle{\nabla^2 \phi_0=0}$.
The two integration constants are determined by the specific type of physical problem under consideration.
One assumes that the source, apart from carrying charge, has also mass.  Thus, $c_1$ is chosen  to be $c_1=-mG_N$ to reproduce the Schwarzschild gravitational potential. The remaining constant, $c_0$, is fixed by the boundary conditions.  We are going to determine $c_0$ in a while.\\
Inserting the solution (\ref{c1}) in (\ref{sol}) we find

\begin{equation}
\phi=c_0 -\frac{mG_N}{r}+2\pi G_Ng^2\, \left[\, \frac{1}{(4\pi r)^2}+  \frac{\Lambda_{QCD}^4}{(8\pi)^2}\, r^2- \frac{\Lambda_{QCD}^2}{16\pi^2}\, \right]\ .
\end{equation}
Now, we shall determine constant $c_0$ in such a way to obtain an AdS space-time 
at large distance \footnote{The alternative choice $c_0=0$ leads to the  presence of a \emph{topological defect} in the origin. In this case the surface of a sphere at large distance is less than $4\pi r^2$.}

\begin{eqnarray}
&& c_0=G_Ng^2 \frac{\Lambda_{QCD}^2}{8\pi}\nonumber\\
&& f(r\to \infty)=1+\frac{G_Ng^2\Lambda_{QCD}^4 }{32\pi}\,r^2\nonumber
\end{eqnarray}

The final form of the metric is

\begin{equation}
 -g_{00}=g_{rr}{}^{-1}= 1 -\frac{2mG_N}{r} +2\pi \,G_Ng^2\left(\, \frac{1}{4\pi r^2}+\frac{\Lambda_{QCD}^4}{32\pi}r^2\,\right) \ .\label{rnads}
\end{equation}

\begin{figure}[h!]
                \begin{center}
                \includegraphics[width=10cm,angle=0]{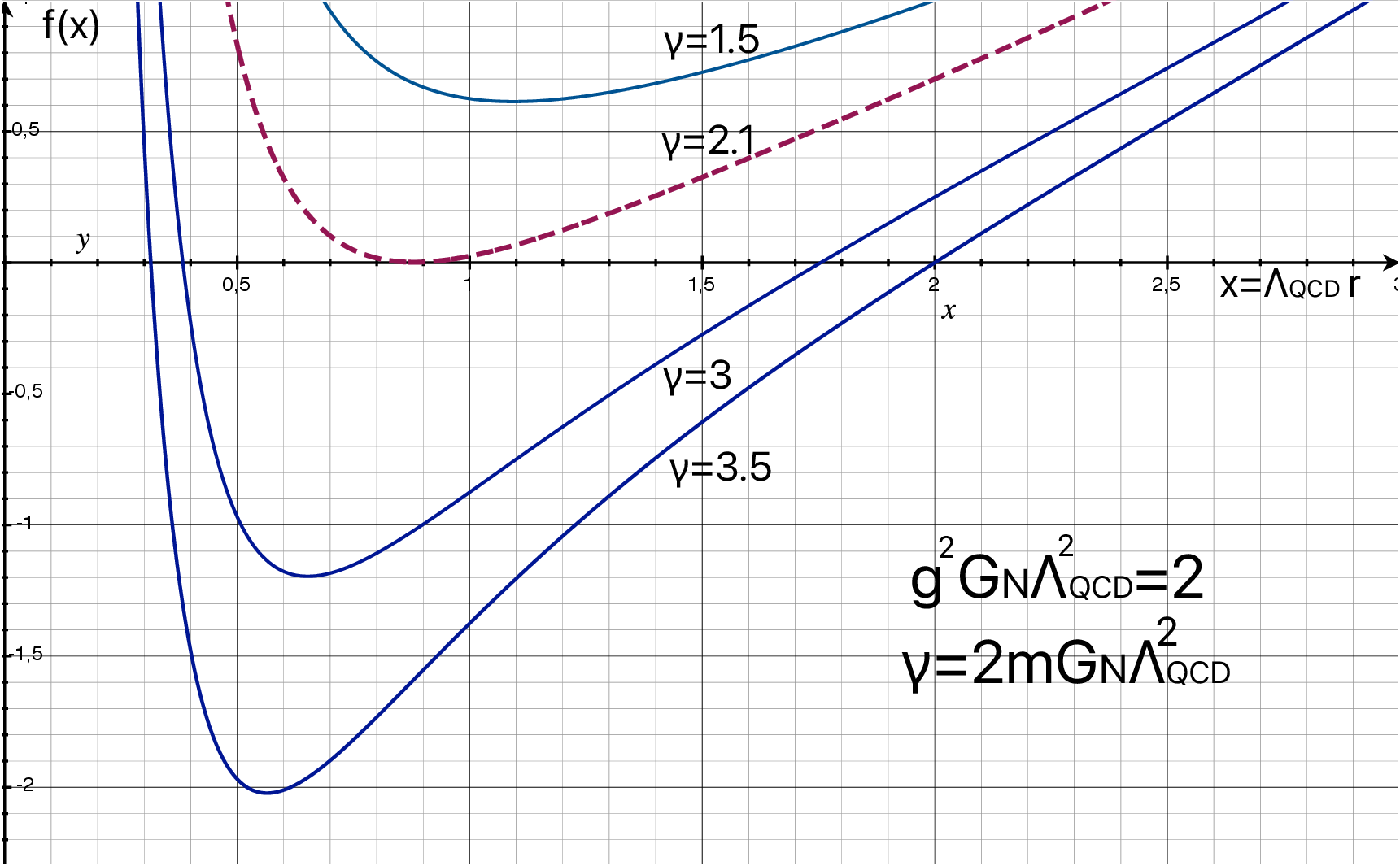}
               \caption{Plot of the metric function for various values of the mass $m$ at fixed $g$ and $\Lambda_{QCD}^2$. For $\gamma> 2.1 $ there two horizons. They merge into an extremal 
                black  hole
                for $\gamma=2.1$ (dashed curve). Finally, if the mass is too small, i.e. $\gamma< 2.1$ there are no horizons. }
            \label{crnlmetric2}
             \end{center}
             \end{figure}
Equation (\ref{rnads}) describes an electrically charged Anti-deSitter geometry. 
\\
We can establish the correspondence between gauge and gravitational parameters as follows:
\begin{eqnarray}
&&\alpha_g\equiv \frac{g^2}{4\pi}=\hbox{gauge coupling constant}\ ,\\
&& \alpha_gG_N\equiv G_s=\hbox{gravi-strong coupling constant }\ ,\\
&&\frac{G_s}{4 }\Lambda_{QCD}^4\equiv \frac{\Lambda}{3}>0=\hbox{effective cosmological constant}\ .\label{cc}
\end{eqnarray}
We remark that, in our approach, the result  is neither a conjectured duality between Yang-Mills theory and higher dimensional gravity
\cite{Maldacena:1997re,Witten:1998zw,Witten:1998qj}, nor
  an "~ad hoc~" identification between gauge and gravitational coupling constants as it is
done in the double copy framework.  
Rather than a conjectured duality, we obtained  an exact connection between a confining gauge model and  4D Einstein gravity.\\
Let us notice that the identification (\ref{cc}) $\Lambda\propto \alpha_s\left(\, \mu\,\right)$ leads to
a "~running cosmological constant~" rather than a fixed one. This property makes it reasonable 
to identify $\Lambda$ as the \emph{pressure} in the thermodynamical description of a CBH.\\
The plot of $f(r)$ for various $m$ is shown in Fig.(\ref{crnlmetric2}).\\

\begin{figure}[h!]
                \begin{center}
                \begin{minipage}[h]{0.4\textwidth}
                \includegraphics[width=7cm,angle=0]{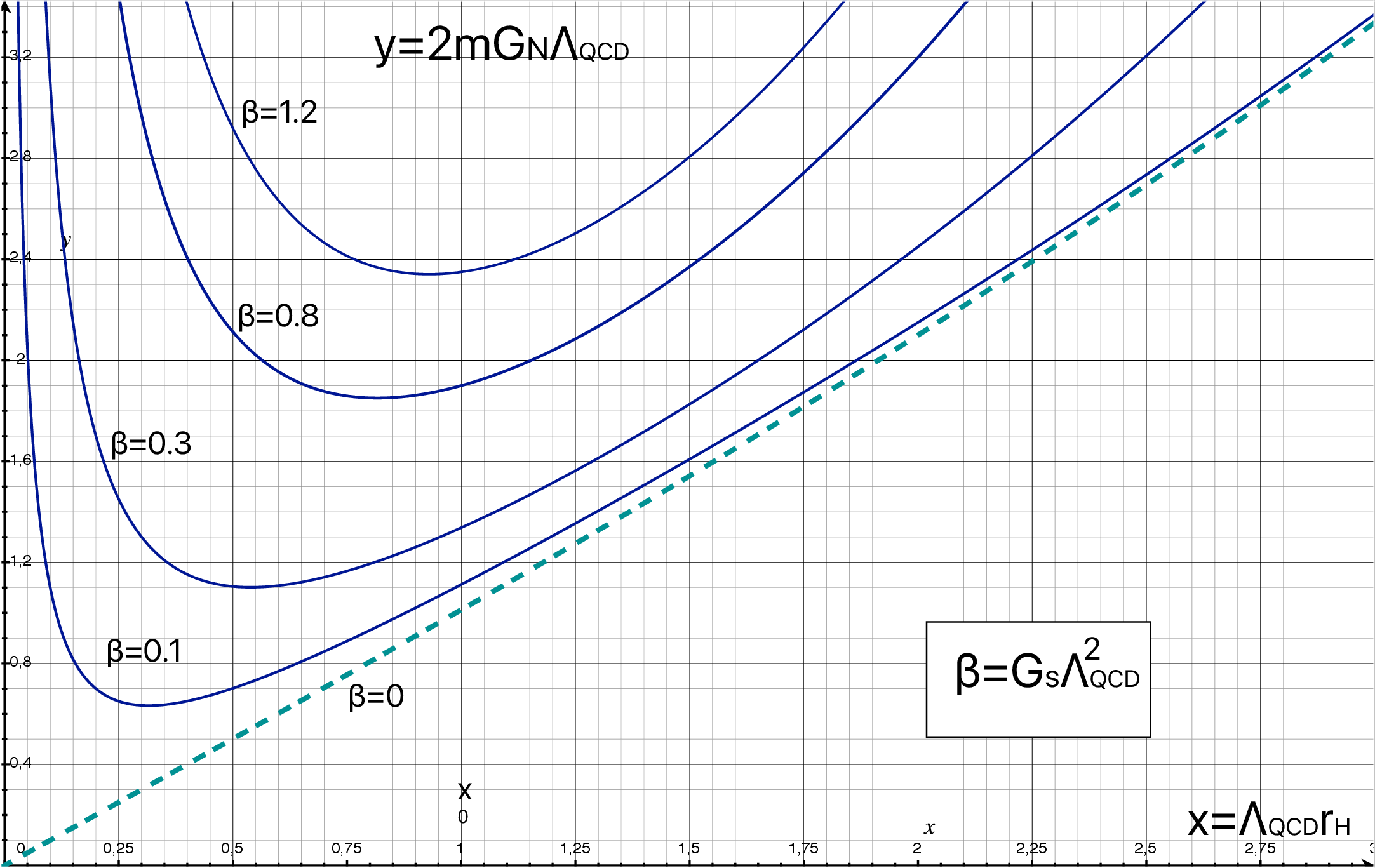}
                \caption{Plot of the horizon equation for various values of $\beta=G_s \Lambda_{QCD}^2$. The minimum of the curves correspond to extremal configuration where the two horizons merge.  }
            \label{crnl2}
             \end{minipage}
             \hfill
\begin{minipage}[h]{0.4\textwidth}
                \includegraphics[width=6.7cm,angle=0]{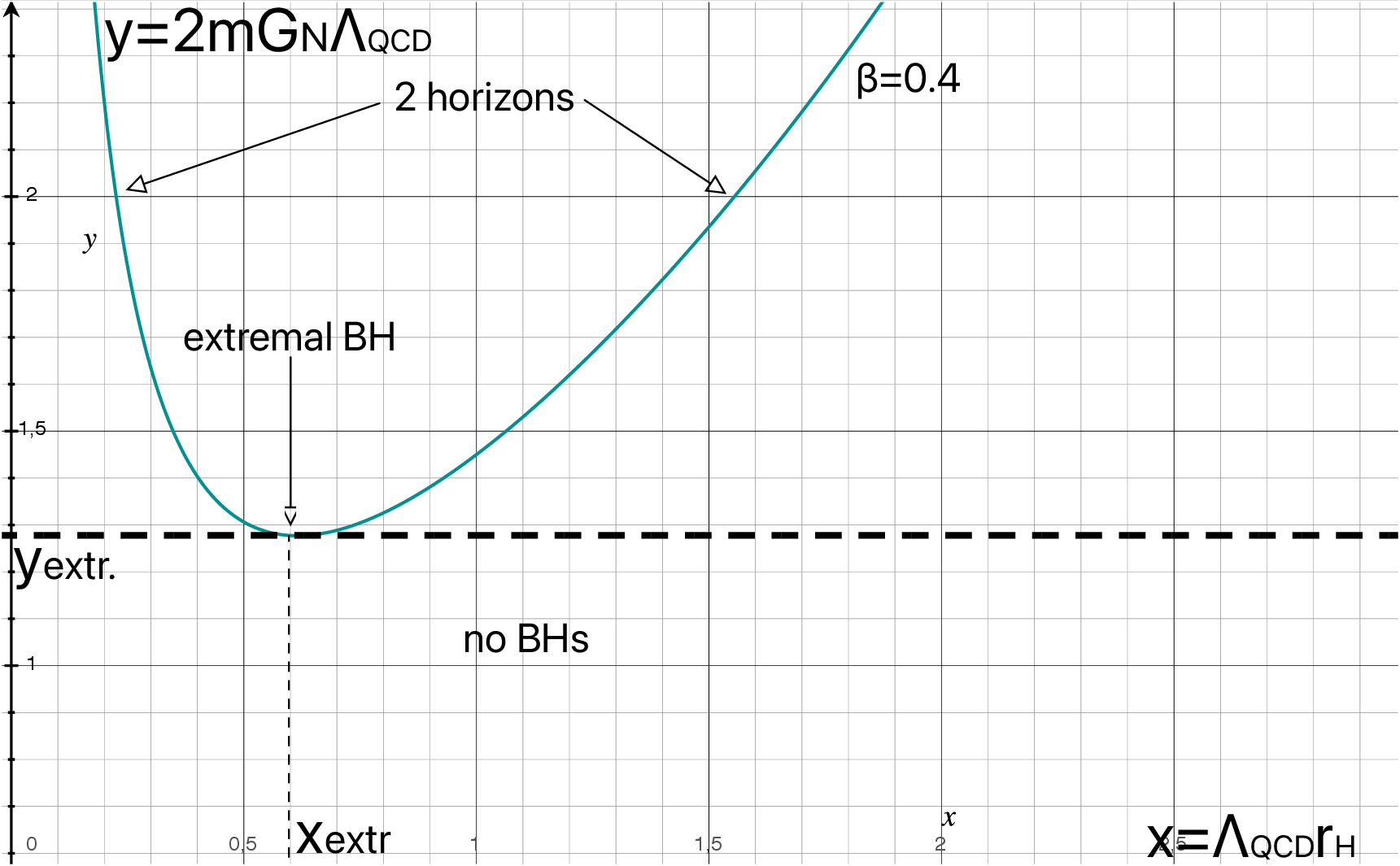}
                \caption{Plot of the horizon equation for  $\beta\equiv G_s \Lambda_{QCD}^2=0.4$. The minimum of the curve corresponds to an extremal BH. For $y>y_{extr}$ there are two horizons; for $y<y_{extr}$ there are no horizons. }
            \label{crnl3}
                          \end{minipage}
             \end{center}
\end{figure}

To investigate the existence of horizons we consider the equation $f(r_H)=0$ and express mass as a function of horizon $r_H$ 

\begin{equation}
2mG_N = r_H + \frac{g^2}{4\pi}G_N r_H \left(\, \frac{1}{ r^2_H}+\frac{\Lambda_{QCD}^4}{8}r^2_H\,\right)
\end{equation}

For parameter $\beta\equiv G_s\kappa^2>0$ there are two horizons which  merge into a degenerate one in the case of an extremal CBH.\\
Let us note that as the cosmological constant is $g$-dependent. Thus, the $g\to 0$ limit of $f(r)$ is not a neutral AdS metric but a simpler Schwarzschild geometry.\\

\section{Thermodynamics analysis of CBH}
\label{termo}
The thermodynamical description  of AdS black holes has been given due attention in several papers,
e.g. \cite{Chamblin:1999tk,Chamblin:1999hg,Caldarelli:1999xj,Smailagic:2012cu}.
In the original work by Hawking and Page 
\cite{Hawking:1982dh}, a phase transition between a  gas of particles and a Schwarzschild-AdS  
black hole was introduced.  The main difficulty in this approach is the proper identification
of (~thermodynamical~) canonical variables. While  fluid temperature can be naturally related to the  Hawking temperature, the identification of other quantities such as pressure, volume, enthalpy, etc., is not so straightforward
\cite{Kastor:2009wy,Dolan:2010ha,Dolan:2011xt,Dolan:2011jm,Dolan:2012jh,Cvetic:2010jb}. 
As an illustration  of this difficulty, consider the cosmological constant.  By
definition is a "~constant~".  However, once it is identified with the fluid pressure, it becomes a function of the temperature and the volume \cite{Kubiznak:2012wp,Gunasekaran:2012dq} . This relation is provided  by the  Van der Waals equation.\\\\
To give a thermodynamical description of the CBH,  we start from the black hole temperature equation
\cite{Kubiznak:2012wp}
\begin{figure}[h!]
                \begin{center}
               \includegraphics[width=10cm,angle=0]{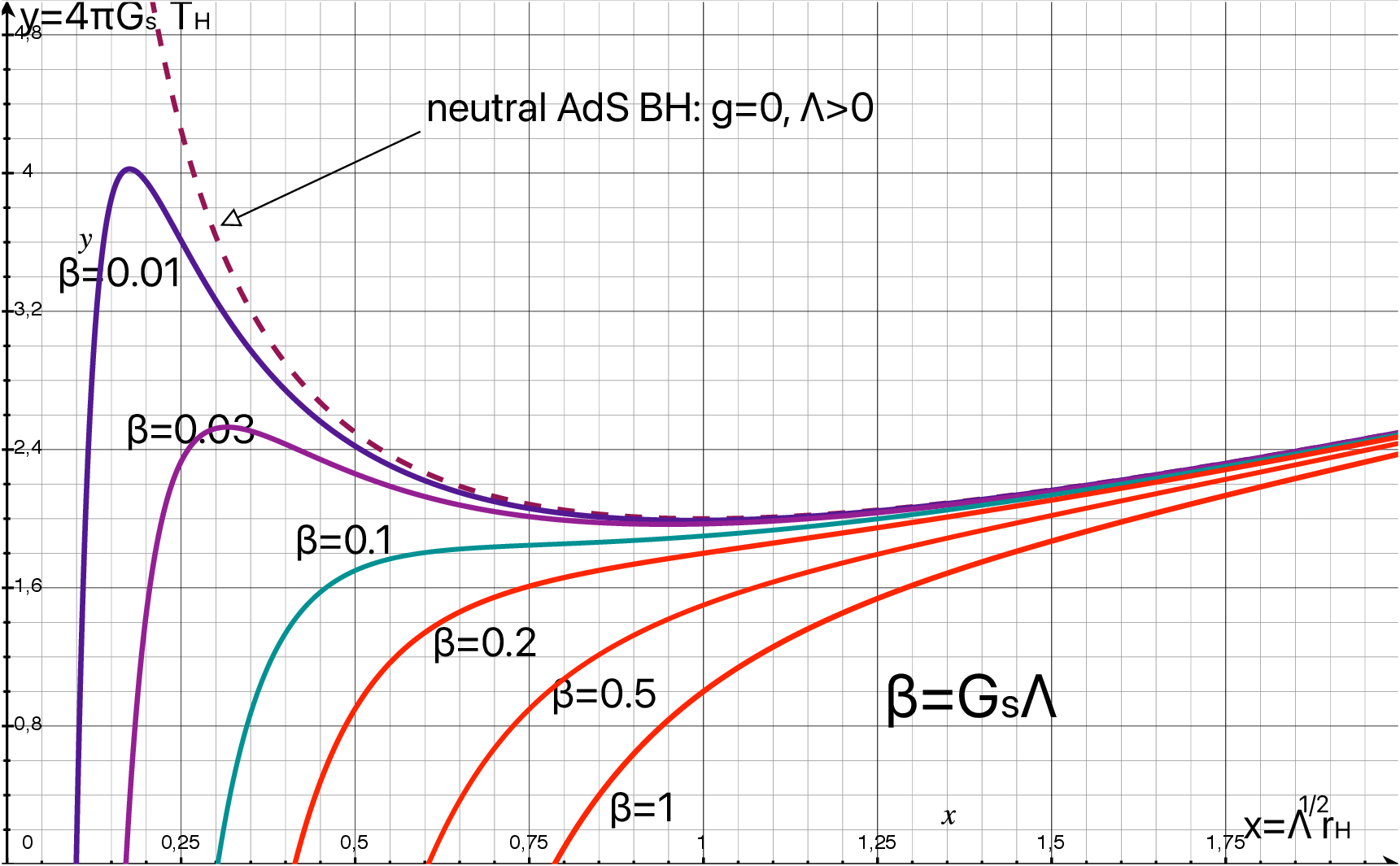}
                \caption{Plot of the horizon temperature as a function of the horizon radius $r_H$. The curves
                are obtained for different values of the parameter $\beta\equiv G_s\Lambda$. 
                The dotted curve represent  the limiting case of a  neutral Schwarzschild AdS black hole.}
            \label{tcorn}
             \end{center}
\end{figure}
\begin{equation}
T_H=\frac{1}{4\pi r_H}\left(\, 1 
-\frac{G_s}{r^2_H}+\Lambda r^2_H\,\right)\label{V}
\end{equation}
and define the pressure and the specific volume as follows:
\begin{eqnarray}
&& P\equiv \frac{\Lambda}{8\pi G_N}\ ,\qquad\hbox{"~pressure~"}\ ,\\
&& v\equiv 2G_N r_H \ ,\qquad\hbox{"~specific volume~"}\ .
\end{eqnarray}
In this way, Eq. (\ref{V}) turns into  a Van der Waals equation ( $\kappa_B\equiv 1$ ):

\begin{equation}
P=\frac{T_H}{v}-\frac{G_N}{2\pi v^2}+\frac{2G_N^3G_s}{\pi v^4}
\end{equation}

\begin{figure}[h!]
                \begin{center}
                \includegraphics[width=10cm,angle=0]{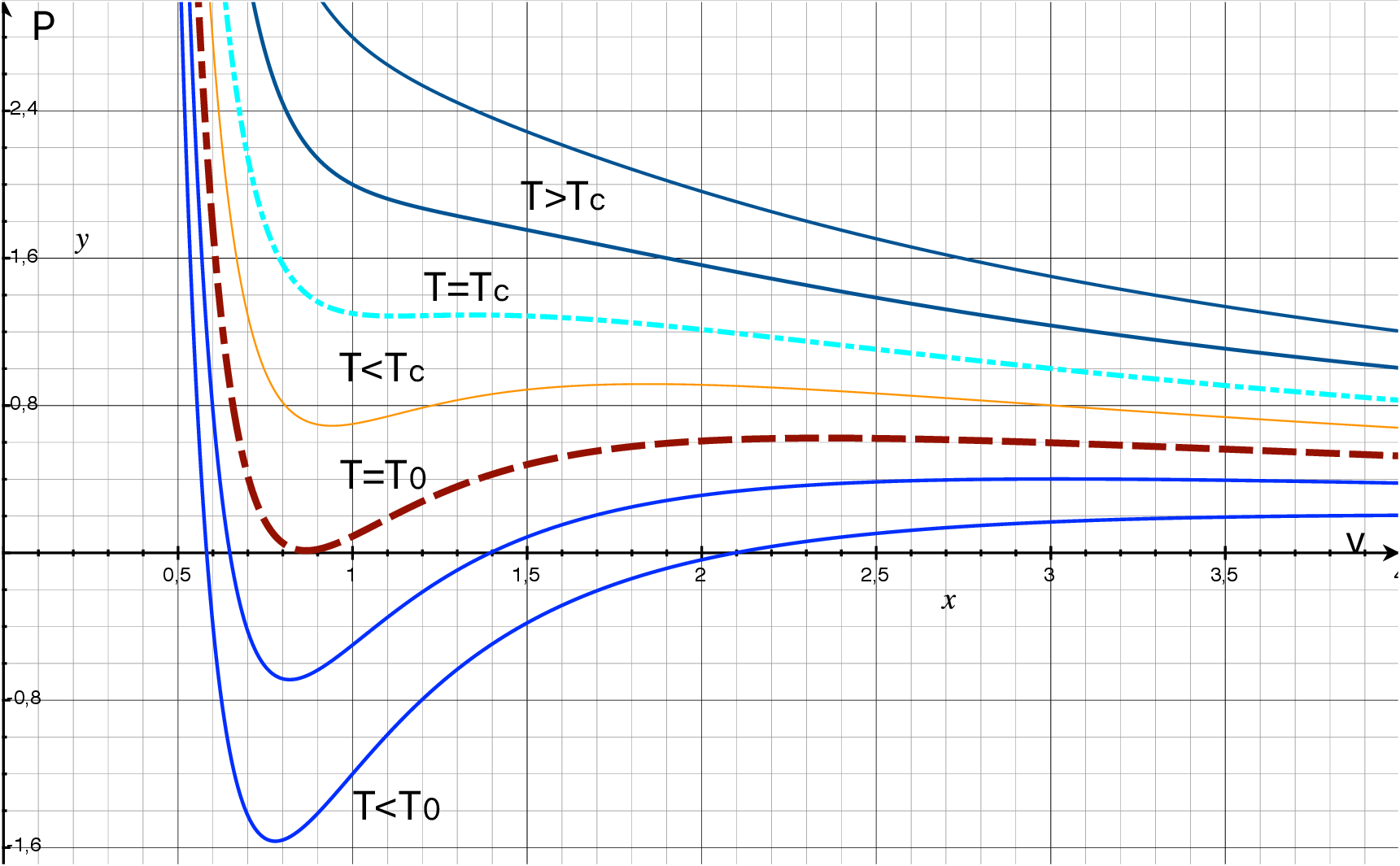}
                \caption{PV diagram for $G_N/2\pi=4$, and $2G_sG_N^2/2\pi=1$. The region $T>T_c$  corresponds to an "~ideal gas~". The dotted line is \emph{critical isotherm} $T=T_c$. The oscillating part of the isotherms for $T< T_c$ must be replaced by isobars according with the  Maxwell's area law. For $T<T_0$ isotherms can show negative pressure intervals. This part of the diagram is
                \emph{unphysical} and is eliminated by applying Maxwell's construction.}
            \label{tcorn}
             \end{center}
            \end{figure}
The phase transition occurs at the critical temperature $T=T_c$, where the isotherm shows an \emph{inflexion point}:

\begin{eqnarray}
&&\frac{\partial P}{\partial v}=0\ ,\\
&&\frac{\partial^2 P}{\partial v^2}=0\ .
\end{eqnarray}

The values of the critical parameters are:

\begin{eqnarray}
&& v_c^2=24G_sG_N^2 
\ ,\label{vc}\\
&& T_c=\frac{\sqrt{6}}{18\pi G_s^{1/2}}
\,\label{tc}\\
&& P_c=\frac{1}{96\pi G_s G_N}
\label{pc}
\end{eqnarray}

Once the CBH is  reformulated as a Van der Waals fluid, then the critical parameters satisfy 
the following relations
\begin{equation}
\frac{P_c v_c}{T_c}=\frac{3}{8}\label{vdw}
\end{equation}

There is a characteristic value, $T_0$, of the temperature  where:

\begin{eqnarray}
&& P=0\ ,\\
&& \frac{\partial P}{\partial v}=0
\end{eqnarray}
\begin{eqnarray}
&& v_0^2=12 G_s G_N^2 
\ ,\\
&& T_0=\frac{\sqrt{3}}{18\pi G_s^{1/2}}\ . 
\end{eqnarray}

For any temperature $T< T_0$ there is  unphysical region where the isotherms show a negative pressure over some interval. To avoid
this problem, isotherms must be replaced by isobars determined by the Maxwell' s area law \cite{Spallucci:2013osa,Spallucci:2013jja}.

\section{Conclusions}
\label{end}
We studied  a non-local Abelian model  leading to a Cornell type potential between static charges.
We also showed that it is dynamically equivalent to Maxwell electrodynamic with an extended source. In
the latter form, it is easier to calculate the contribution from the interaction term $J^\mu A_\mu$ to the
full energy momentum tensor. The resulting $T^{\mu\nu}$
is used as the source for solving the Einstein equations.\\
 A  static solution for the metric  is easily obtained
in the Kerr-Schild gauge because Einstein equations reduce to the Poisson equation in flat
space (\ref{Poi}). 
Written in the spherical gauge, the exact solution is  a charged Anti deSitter metric.
We consider this result  as a novel and important way to establish an exact relationship between the Cornell potential and the charged AdS black hole. 
\\
In our construction the cosmological constant is related to the strong running coupling constant 
$\alpha_s(\mu^2)$ (\ref{cc}). Thus, it allows natural way to identify  $\Lambda$ 
 as the pressure of a "~Cornell photon fluid~" at the Hawking temperature.
Van der Waals description of the CBH is characterised by gas-liquid phase transition. We determined the critical parameters (\ref{tc}),(\ref{pc}),(\ref{vc}) for such a transition to occur. 
\\
\section*{Appendix: Interaction energy for distributed charge.}
In the literature the interaction term
\begin{equation}
S_{int}= e\int d^4x \,\sqrt{-g}\,g_{\mu\nu}\, J^\mu A^\nu \ ,\label{sint}
\end{equation}
is not taken into account in calculating $T^{\mu\nu}$. This is a legitimate procedure  for a point-like charge, as the current density $J^\mu$ is zero everywhere except along
the particle world-line. But here the field itself is not defined. Accordingly, $T^{\mu\nu}$ is derived
from the variation of the free field term alone.\\
When a source is not point-like, the interaction term has to be properly accounted for.  Despite the appearance, it is important to remark that the variation of
$S_{int}$  has to be done \emph{only} with respect to the metric tensor $g^{\mu\nu}$ and not on the metric determinant $g$. To clarify this claim,
let us consider  a point-like charge. Its associated current reads
\begin{equation}
J^\mu = \frac{1}{\sqrt{-g}}\delta^{4}\left[\, x - x(\tau)\,\right] \dot{x}^\mu\ ,
\end{equation}
and the action is

\begin{equation}
S_{int}= e\int d^4x\, \delta^{4}\left[\, x - x(\tau)\,\right]\, g_{\mu\nu} \dot{x}^\mu A^\nu \ ,\label{sint0}
\end{equation}
where the Dirac delta is defined by the normalization condition

\begin{equation}
\int d^4x \,\delta^4(x)=1\ ,\quad \hbox{or}\qquad \int d^4x\sqrt{-g} \,\frac{1}{\sqrt{-g}}\delta^4(x)=1\ . \label{norm}
\end{equation}

In  the case of non-point like charges the Dirac delta is replaced by a charge distribution $\rho(x)$ normalized by the same
condition (\ref{norm}). The current density becomes
\begin{equation}
J^\mu =\frac{1}{\sqrt{-g}}\,\rho(x) \, u^\mu(x)\ ,\qquad u^\mu u_\mu =-1\ .
\end{equation}
It follows that the variation of the action is now given by
\begin{equation}
\delta S_{int}= e\int d^4x \,\delta g_{\mu\nu}\, \rho(x)\, u^\mu A^\nu\equiv  \int d^4x \sqrt{-g}\delta g_{\mu\nu}\, T^{\mu\nu}{}_{int}\ ,
\end{equation}
and the contribution of the interaction term to the energy-momentum tensor is 

\begin{equation}
T^{\mu\nu}{}_{int}=\frac{1}{2} J^{(\,\mu}A^{\nu\,)}\ .\label{last}
\end{equation}
In (\ref{last}) world indices are symmetrized.

\end{document}